# OCCURRENCE OF A CYBER SECURITY ECO-SYSTEM: A NATURE ORIENTED PROJECT AND EVALUATION OF AN INDIRECT SOCIAL EXPERIMENT

**Utku Kose**

*Suleyman Demirel University, Faculty of Engineering, Computer Engineering*

utkukose@sdu.edu.tr

**Abstract**

Because of today's technological developments and the influence of digital systems into every aspect of our lives, importance of cyber security improves more and more day-by-day. Projects, educational processes and seminars realized for this aim create and improve awareness among individuals and provide useful tools for growing equipped generations. The aim of this study is to focus on a cyber security eco-system, which was self-occurred within the interactive educational environment designed under the scope of TÜBİTAK 4004 Nature Education and Science Schools Projects (with the name of A Cyber Security Adventure) with the use of important technologies such as virtual reality, augmented reality, and artificial intelligence. The eco-system occurred within the interactive educational process where high school students took place caused both students and the project team to experience an indirect social experiment environment. In this sense, it is thought that the findings and comments presented in the study will give important ideas to everyone involved in cyber security education, life-long learning processes, and the technology use in software oriented educational tools.

## 1. Introduction

After rising of the computer technology, value of the information has increased rapidly because of almost infinite opportunities of processing information through digital systems and obtaining effective solutions for real world problems. Moreover, development of innovative communication technologies such as Internet has made it easier to carry information point to point and ensure instant, practical information processing environment spread all around the world. As a result of that revolutionary location change of information (from minds, papers to the digital systems) and intense use of digital systems for every daily life operations, the world has started to use the term: 'data' widely instead of the concept of information (Loukides, 2011; Mehlhorn, 2013). Although creating, processing, or transferring data have changed the humankind's life totally and event resulted to transformation of the society to an 'informatics society', storing the information in the form of data has caused appearance of security issues. Because of alternative ways to 'hack' digital systems and possibility of losing the data because of environmental factors, keeping the data secure has become more and more important day-by-day (Hu, 2016; Pfleeger, & Pfleeger, 2002; Stallings et al., 2012). Today, there are many different approaches, methods, and techniques to hack the data / digital systems or build somehow security against dangerous and / or malevolent scenarios. Covered by the topic of 'cyber security', design and development of different solutions always has a top importance because of unstoppable new technological developments changing the way of threatening or ensuring cyber security (Baig et al., 2017; Choo, 2011; Dawson, & Thomson, 2018). However, since many malicious attacks include also person / user tricking phases (social engineering) and complete their process successfully when they are applied for untrained people / users, cyber security education has become a necessity for today's individuals.

Because of need for trained and 'aware' people / users against cyber security issues, there is a remarkable effort to conduct courses, educational seminars or different-sized projects in all over the world. In the USA, there is already a long-time-performed serious works on how to process successful cyber security education, design effective curriculums, and rise effective workforce in this manner (Conklin et al., 2014; Ivy et al., 2019; McGettrick et al., 2014; Newhouse et al., 2017; Shoemaker et al., 2018). As the heart of the technology, Japan has already been discussing effective ways for successful cyber security education (Beuran et al., 2016). There are already some reported studies from India, and Malaysia (of Far East), in the context of ensuring better cyber education and discussing current state of awareness regarding especially cyber security and laws (Mehta, & Singh, 2013; Muniandy et al., 2017; Saxena et al., 2016). In Europe, dominant locations such as UK and Germany have reported studies regarding cyber security education, awareness, and organization of workforces (Bada, & Nurse, 2019; Brittan et al., 2018; Hoffman et al., 2011). As including more of Europe, and other regions such as Africa or South America, there have already been active studies regarding cyber education and cyber security awareness, considering even cyber wars (Catota et al., 2019; Knox et al., 2018; Lejaka et al., 2019; Parker, & Brown, 2018; Tokola et al., 2019; Von Solms, & Von Solms, 2014). As it has already been reported that technology supported competitions have important potential for alternative outcomes in cyber security education (Buchler et al., 2018; Cheung et al., 2012; Pusey et al., 2016), research efforts have been directed to design of alternative activities where educators and learners may take active part by experiencing game-based, real component-based (i.e. robotics), virtual-lab supported, or team oriented (i.e. Capture the Flag) processes (Cheung et al., 2011; Chothia, & Novakovic, 2015; Giannakas et al., 2015; Jin et al., 2018; Khoo, 2019; Olano et al., 2014; Rand et al., 2018; Werther et al., 2011; Willems et al., 2011). At this point, importance of managing educational processes (whether they are traditional or innovative) has been often emphasized and even learning-teaching styles, emotional levels, or effects of digital / mass media have been evaluated for seeking better educational outcomes (Buckingham, 2015; Debeş, & Öznacar, 2018; Graziose et al., 2016; Johnson, 2015; Khalid et al., 2017; Öznacar, & Dagli, 2016; Öznacar, & Dericioğlu, 2017; Öznacar et al., 2017; Öznacar et al., 2018; Parrish, 2015; Van der Sandt, & O'Brien, 2017). It seems that every educational process may require different managerial requirements and cause unexpected situations that can contribute to the outcomes differently. As one of the best ways to run today's cyber education processes seems to be conducting them within projects, a unique project called as 'A Cyber Security Adventure' was performed in Isparta, Turkey, during second half of 2018 and first months of 2019 (Kose, 2018). Except from the other tasks such as project process organization, development of software, and past-activity findings analysis, main educational activities of the project were done between 27 August 2018 and 9 September 2018, by including two stages: face-to-face education, and nature camp, respectively. In addition to the findings reported in detail by Kose (2018), the project has also included many different findings, generally occurred from unexpected states caused by dynamic nature of educational flow.

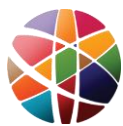

Based on the explanations so far, the aim of this study is to focus on a cyber security eco-system, which was self-occurred within an interactive educational environment designed under the scope of TÜBİTAK 4004 Nature Education and Science Schools Projects. With the use of important technologies such as virtual reality, augmented reality, and artificial intelligence, the related interactive educational environment was realized under a unique project titled as 'A Cyber Security Adventure' (with also an acronym: *BSGM*, as derived from the Turkish title: *'Bir Siber Güvenlik Macerası'*). Except from the findings-results obtained with the planned evaluation methods (as reported in also Kose, 2018), active project environment has caused to creation of an unexpected eco-system. The eco-system occurred within the interactive educational process where high school students took place caused both students and the project team to experience an indirect social experiment environment. Although findings from planned evaluation methods of different interactive activities or project organizations may have great importance in terms of literature, some hidden details and unexpected experiences-outcomes can give more information about benefits of the formed educational environments and open minds for further, alternative research studies. In this sense, it is thought that the findings and comments presented in the study will give important ideas to everyone involved in cyber security education, life-long learning processes, and the technology use in software oriented educational tools.

Considering the topic and the aim of the study, the remaining content is organized as follows: The next section briefly gives essential information about the realized project. In this way, the readers are enabled to have enough information about why that project was designed and how it was organized accordingly. After that section, the third section explains some remarkable findings obtained thanks to the project. Although these findings may be subject to other studies (as the project has many different perspectives to write down different reports, papers…etc.), it is also important to inform the target readers about the findings briefly. Next, the fourth section focuses on the unexpected eco-system and some observations associated with that eco-system. That fourth section has become the main section explaining the exact research achievements pointed in this study. After the fourth section, the content is ended with the final, fifth section including discussions regarding conclusions and some thought future works.

## 2. A Project in the Nature: A Cyber Security Adventure

In Turkey, TÜBİTAK (The Scientific and Technological Research Council of Turkey) opens project calls in the context of different scopes. As one of these calls, the call numbered with 4004 is associated with educational projects aiming to combine both scientific perspectives and interactive activities done at the nature. In this sense, the project: 'A Cyber Security Adventure' (with also an acronym: *BSGM*, as derived from the Turkish title: *'Bir Siber Güvenlik Macerası'*) was accepted to be funded by TÜBİTAK during 6 months period between 27 July 2018 and 27 January 2019 (Figure 1 shows both TÜBİTAK call logo and the project logo). As supported with a total of 80.097 Turkish Liras (around 14.580 USD in August 2019) funding, the project aimed to improve high school students' awareness, knowledge, abilities towards cyber security, thanks to different educational activities (Kose, 2018).

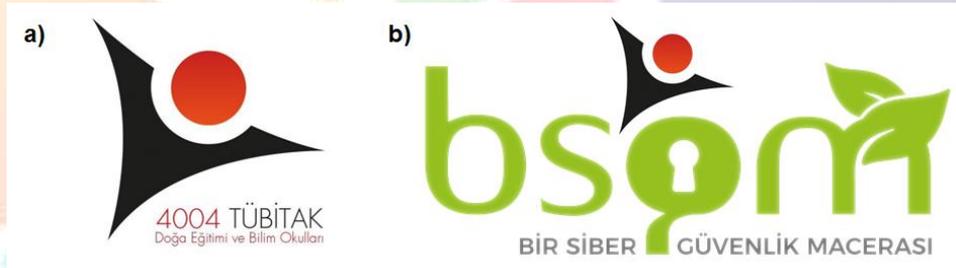

**Figure 16: a) TÜBİTAK 4004 call logo – b) A Cyber Security Adventure (BSGM) project logo (Kose, 2018).**

As the project lead by Assistant Prof. Dr. Utku Kose (from Dept. of Computer Engineering at Suleyman Demirel University, Turkey), call for students and selection of them were done collaboratively with also Isparta Directorate for the Ministry of Education. Because the city of Isparta in Turkey was chosen as the project location, a total of 40 high school students from 10 high schools (with 4-student group of each school) enrolled in the project activities. As it was indicated in the first section, main educational activities of the project were done during 14-day period between 27 August 2018 and 9 September 2018. Between 27 August 2018 and 31 August 2018, 40 students experienced face-to-face education period done by academicians from the Dept. of Computer Engineering at Suleyman Demirel University. Except from training regarding the educational software-applications that will be used later, the following subjects were considered during the face-to-face education period (Kose, 2018):

- Essentials of computer and communication technologies,
- Essential components of today's cyber world,
- The concepts of system, and algorithm,
- Importance of cyber security and awareness regarding it,
- Importance of being ethical (white-hat) hacker,

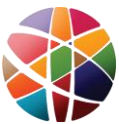

- Theoretical and practical background for cyber-attack,
- Theoretical and practical background for cyber-defense,
- Importance of social engineering and ways to be defended from it,
- Importance of cyber bullying ways to be defended from it.

Following to the face-to-face education period, a 9-day nature camp was held in Egirdir county of Isparta. 40 students from 10 different schools were grouped into 10 teams (as each school forming a team) for the nature camp. The nature camp was done in the natural environment of the Green Island District (called as Island in Egirdir) as the whole process activities were tracked by the project staff. Each team was also directed by one guide attended to it (Kose, 2018). Figure 2 represents a photo of Green Island District of Egirdir and some photos from activity times.

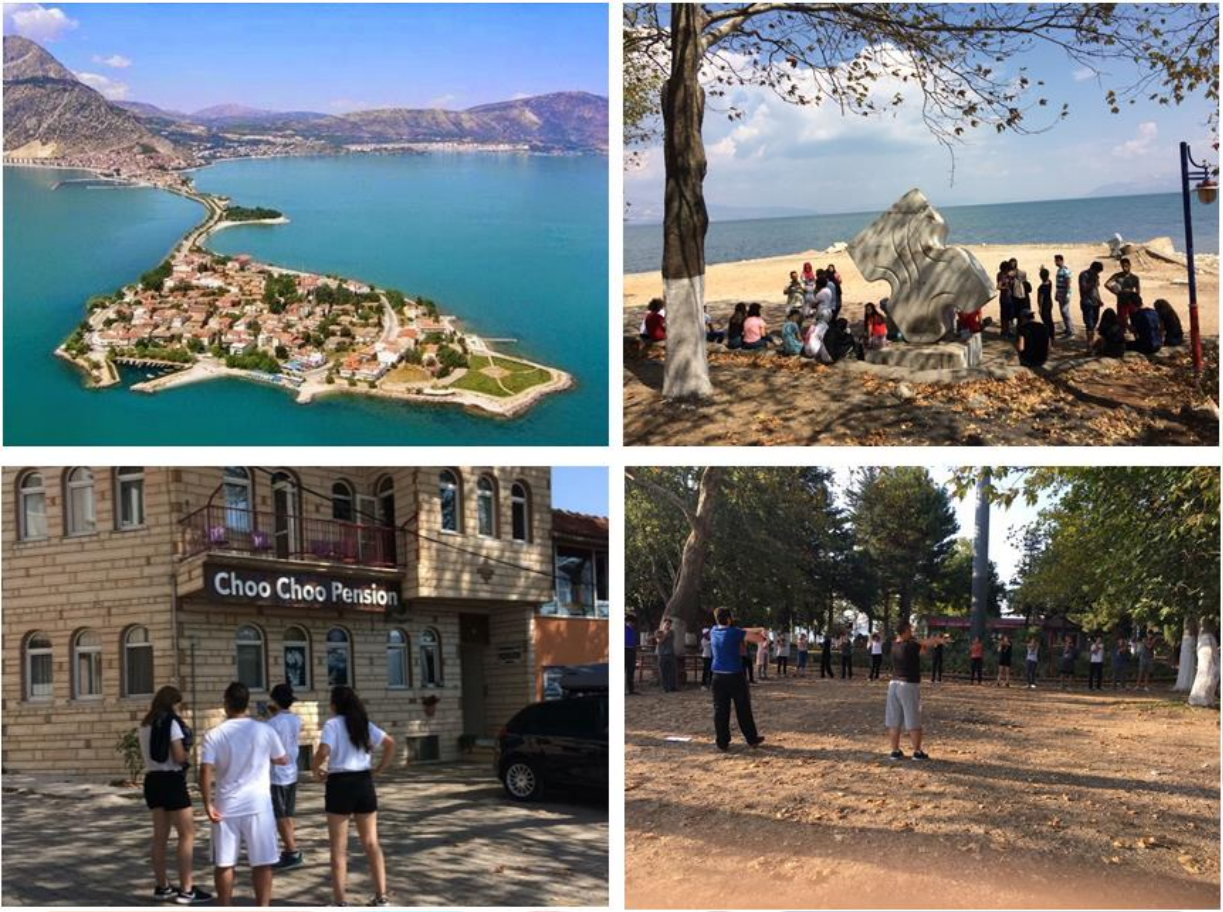

**Figure 17: Green Island District of Egirdir and some photos from project activity times, respectively.**

The interactivity ensured for better educational outcomes during the nature camp were done with the support by technology and educational software use. The following sub-section provides brief information regarding that.

**2.1. Technology and Educational Software Use in the Project**

A Cyber Security Adventure consisted of using some technological approaches and educational software systems for ensuring a high interactivity. In this context, a virtual scenario asking the students-teams to find the cyber pirate among 5 suspected virtual characters were run on the background. In detail, all the nature camp was a combination of dynamically changing sub-stories supporting the main scenario so that the cyber pirate can be determined step-by-step, according to each successfully solved sub-story (Kose, 2018). The following technology and educational software use approaches supported the whole scenario flow and a high level interactivity was achieved in this way (Kose, 2018):

- **Mobile Device Use:** Nature camp included intense use of mobile devices in order to interact with the real world and ensure a bridge between the real world and the virtual scenario. In this context, each team had one tablet device and everybody allowed using their own mobile devices (i.e. smart phones or other mobile devices).
- **Augmented Reality:** As a combination of both real and virtual world (Furht, 2011; Kose, 2015), Augmented Reality is often used for software systems aiming to run high level interactivity. In the nature camp, it was aimed to run Augmented Reality for collecting virtual materials spread in the real world and interact with QR codes or specially designed signs (located over i.e. buildings, trees, or other real components) for activating sub-scenarios or solving puzzles.

- **Artificial Intelligence:** As one the most effective technology of today's world (Rabelo et al., 2018; Russell, & Norvig, 2016), and widely used for developing intelligent educational software systems, Artificial Intelligence was used for designing an adaptive infrastructure, which is directing each team through the virtual scenario. As inspiring from Role Playing Games (RPG), each team consisted of 5 attributes: Speed, Team Spirit, Ability, Energy, and Resource. According to different solutions by students of the teams (or in case of unsolved problems) for sub-scenarios / activities, these attributes were adaptively scored by the intelligent system running on the background. Also, thanks to an intelligent combinatorial optimization approach (Du, & Pardalos, 2013; Siarry et al., 2016), the best sub-scenario flow for each team was determined by the intelligent system, from many different possible sub-scenarios designed before.
- **Software Systems / Applications:** During nature camp, the teams were wanted to use some developed software systems / mobile applications, in order to solve puzzles, enroll in competitions, gain points individually, and eventually go through the sub-scenarios. In order to benefit from their knowledge and abilities gained during face-to-face education period, students used software systems / applications called as 'Attacking Machine', 'Defensing Machine', 'Tracker', 'Reader', and 'Code Breaking Machine'.
- **Social Media:** As an essential technological component for the nature camp, social media was used for supporting the interactivity and communication among all people including students, and even project staff. Thanks to groups created over WhatsApp, instant problems to be solved in some allowed time period were announced and other social media channels were used for better interaction and information share.

**2.2. General Educational Activities in the Project**

Considering the technological components expressed briefly under the previous sub-section, general educational activities in the project were organized in the context of the following activity scopes (Kose, 2018):

- Theoretical and applied lectures (during 5-day face-to-face education period),
- Group studies (during 9-day nature camp period),
- Project based studies (during 9-day nature camp period),
- Drama studies (during 9-day nature camp period),
- Analytical and logical problem-solving (during 9-day nature camp period),
- Sportive competitions (during 9-day nature camp period),
- Individual problem solving (during 9-day nature camp period),
- Problems requiring social interaction (during 9-day nature camp period),
- Software / application use (during 9-day nature camp period).

**3. Findings from the Project**

Statistically, a composite approach was applied for obtaining findings regarding students, and their activities during the project period. For getting information about students' perception and attitudes against cyber security, a pre-test and post-test approach was applied over a survey. Also, interviews and observations were done for getting more information about students. Additionally, some more technical tests for understanding more about effectiveness of the technological components and their usability levels were applied. Furthermore, some findings from project staff (especially from guides) were obtained for additional analyses. All these findings have been already reported by Kose (2018). But considering the whole findings from the composite approach for students and some technical analyses, the following points can be expressed briefly (Kose, 2018):

- Students' perception and attitude levels changed positively after the project (That affected also some people from project staff).
- Students enjoyed the whole process and wanted to experience similar processes again (That is same for also project staff).
- Students were highly motivated for the issues of cyber security, after the project period.
- Students, who has some past knowledge and ability regarding cyber security and computer use, contributed to their team greatly.
- Teams, who created more team synergy, has achieved better during nature camp.
- Teams were generally aware of using their team attributes against solving sub-scenarios / problems / puzzles.
- The project contributed to cyber security awareness and gaining knowledge-ability against current state of the digital world.
- Interactivity and the natural environment contributed positive outcomes greatly.
- It was shown that a combination of nature (real world) and technology (digital-cyber world) can contribute to educational processes better.
- The project took attention of national and international groups enrolled in cyber security, technological developments, and the digital world.
- Especially nature camp has caused an unexpected cyber security eco-system of (as the main topic of this study, explained under the fourth section).
- A remarkable number of students improved their self-confidence, leadership, and academic level, thanks to the project.
- The project contributed positively to some students' future career ideas. These students generally motivated about choosing their career to move within cyber security, computer (software and/or hardware) technology, or communication technology.
- Software systems / applications were generally usable and efficient enough in terms of enabling students and project staff to focus on only task and perform their tasks within desired time periods.
- The Artificial Intelligence infrastructure was successful enough to run an adaptive scenario flow, according to dynamically changing parameters of the nature camp and teams.

## 4. Occurrence of a Cyber Security Eco-system in the Project

Especially nature camp period of the project included a wide interaction among students, project staff. Additionally, that interaction was somehow adaptively controlled / directed by the intelligent infrastructure and the running software systems / applications. Although the virtual scenario included lots of sub-scenarios and alternative ways to complete it (even some sub-scenarios were not activated by the teams), it was at least having a certain plan designed before. On the other hand, there were some instant touches by the project leader: Dr. Kose, in order to ensure a dynamic scenario flow. However, the nature camp faced also an unexpected formation of a cyber security eco-system, because of competition and ambition increased day-by-day. That situation has affected even some of project staff and eventually the nature camp turned into a also a social experiment. The related eco-system and the associated indirect social experiment are all remarkable to be analyzed, as valuable, additional findings.

The indirectly appeared cyber security eco-system had the following components with their potentials in that eco-system:

- **Teams / Students:** Because of high competition level, teams / students were important to eliminate each other, by considering every type of fake activities or counter-activities.
- **Guides:** Guides were responsible for controlling students and directing them to solutions if they find the problem / sub-scenario difficult to solve. But because they were informed about only general framework of the scenario, they knew nothing about how to follow true ways of the scenario and how much their team can gain from each sub-scenario / problem, which were even created by the project leader sometimes. So, guides had always potential to be hooked in competition and ambition.
- **Nature / Environment:** Components for especially Augmented Reality, and software interactions (so that interaction with the nature / environment) were located over real components such as buildings, trees, signs, or even other things such as cars, daily life materials…etc. Because of that, they were potential tools within an eco-system, which can be used by everybody for tricking others and even Artificial Intelligence infrastructure.
- **Artificial Intelligence / Software Systems / Applications:** All software oriented components including also Artificial Intelligence infrastructure had the potential of to be tricked because of their digital nature. That is exactly an important actor of a typical cyber security eco-system since it is associated with the digital world.
- **People Out of the Project:** In some sub-scenarios, students' tasks included interaction with people out of the project. These people were generally shop owners or somebody chosen from near environment. They had been informed before about the sub-scenarios and about their responses-roles against students. But they had the potential to be tricked and even people having no idea about the project were potential individuals, so that students can mistakenly think that they know something about the scenario.

In the context of the mentioned components, competitive environment has caused formation of an eco-system, where the following events occurred:

- Some teams having members, who know computer programming, tried to hack the software systems / applications.
- Students started to create fake QR codes for locating them over real components, so that they can trick other teams and even guides (Figure 3).
- Over social media, students started to send fake messages and announcements to trick other teams and directing them to distant locations of the environment so that they cannot complete new, sudden tasks in time.
- Students started to arrange fake people to trick other teams / students about solutions of sub-scenarios.
- Students started to talk with other teams' guides, in order to trick them.
- Some teams even developed fake software or virtual materials so that other teams can use them (by wasting their time) and collecting unnecessary virtual materials.
- There were some soft arguing / discussions between students of different teams.
- Instead of their team members (students), some guides started to give more efforts in order to be successful in physical activities and even argued with some other guides sometimes.
- In some activities requiring knowledge, it was reported that some guides helped their team.
- Some two-three teams become allies against certain teams during physical activities or within 'Attacking Machine' or 'Defending Machine' applications.
- As general, social engineering oriented attacks become popular among teams.
- Teams started to give more consideration to defending themselves, in addition to their attacking activities.
- As similar in real world, some teams acted like cyber pirates to eliminate other teams and some other teams also acted as ethical hackers to save the environment from such teams. Thankfully, teams acting like cyber pirates turned back to the ethical side later.

As it can be understood, an indirect social experiment rose during the nature camp and caused teams and project staff to experience a real world like cyber security eco-system, where members can change side, become allies, run different methods for tricking purpose, and try to defend themselves against possible attacks.

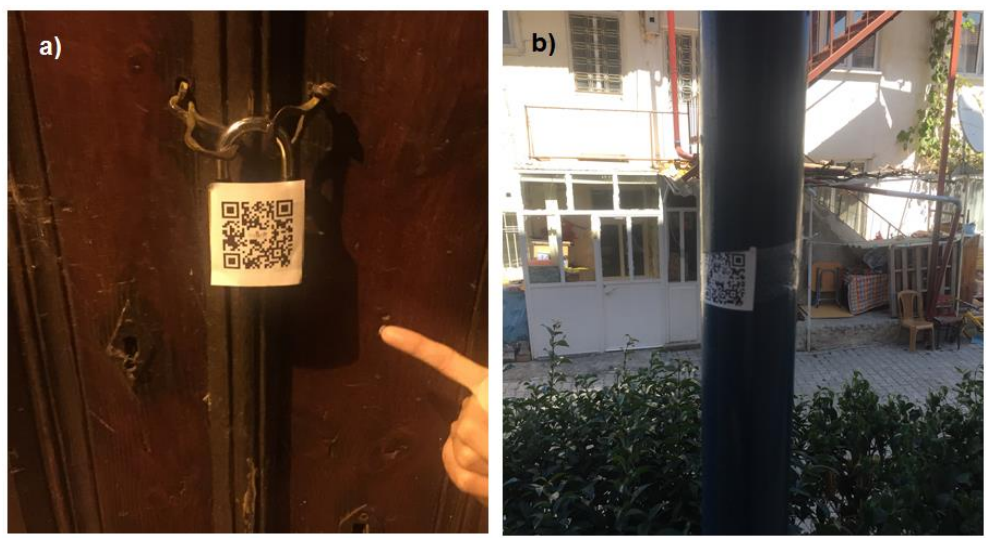

**Figure 18: a) A real QR code located by the project staff – b) A fake QR code created by the student teams.**

After seeing existence of the related eco-system and social experiment flow, the project leader: Dr. Kose decided to gather the related observation notes as the findings just expressed under the previous paragraphs. Additionally, he also analyzed some numerical findings regarding tricking activities done by each team. Table 1 provides total number of tricking activities by each team for each nature camp day, by total number of different tricking activities done as using fake QR-materials, sending fake messages, applying social engineering, direct attack to team(s), direct defense against team(s), attacking with ally / allies, and defense with ally / allies. Names of the teams during the nature camp were used in the Table 1 accordingly. On the other hand, tricking intensity levels has shown in table cells, by using four colors; white: zero, blue: light, yellow: medium, and red: high.

**Table 8: Tricking activities by each team during the nature camp period.**

| Team | Tricking activities (as numbers and color)[a] | | | | | | | | |
|---|---|---|---|---|---|---|---|---|---|
| | Day 1 | Day 2 | Day 3 | Day 4 | Day 5 | Day 6 | Day 7 | Day 8 | Day 9 |
| Phoenix | 0 | 0 | 2 | 1 | 1 | 2 | 3 | 3 | 0 |
| Full-moon | 0 | 0 | 3 | 4 | 6 | 8 | 10 | 5 | 1 |
| Tiger | 0 | 0 | 0 | 2 | 8 | 8 | 3 | 1 | 0 |
| Lightning | 0 | 2 | 4 | 8 | 12 | 7 | 9 | 7 | 1 |
| Wind | 0 | 0 | 2 | 3 | 5 | 5 | 6 | 3 | 2 |
| Mars | 0 | 1 | 2 | 3 | 3 | 5 | 5 | 5 | 2 |
| Pioneer | 0 | 1 | 3 | 4 | 9 | 6 | 3 | 2 | 0 |
| West | 0 | 0 | 2 | 1 | 3 | 3 | 4 | 3 | 0 |
| Cheetah | 0 | 0 | 0 | 0 | 0 | 1 | 2 | 1 | 0 |
| Ghost | 0 | 0 | 1 | 2 | 3 | 5 | 3 | 0 | 0 |

[a]Tricking intensity is shown with four colors; white: zero, blue: light, yellow: medium, and red: high.

As it can be seen from Table 1, more competition and ambition increasing day-by-day caused most of teams to use tricking ways. Although some teams were more robust against tricking, tricking levels for all teams show existence of an eco-system. The eco-system shows high intensity during especially middle of the nature camp and starts to disappear towards the end of the camp. It is remarkable that both students and project staff enjoyed that indirect social experiment and expressed their positive ideas about that suddenly appeared situation, after the end of the nature camp. It is certain that this occurrence of the eco-system and the indirect social experiment has contributed to the project for more positive educational outcomes.

**5. Conclusions and Future Work**

In this study, an unexpected occurrence of a cyber security eco-system in the context of the TÜBİTAK 4004 project: 'A Cyber Security Adventure' was discussed generally. That eco-system has caused also rise of an indirect social experiment, which has contributed to the project outcomes in a positive manner. As including a total of 40 students directed by 10 guides, and also other project staff, almost everybody enrolled in the nature camp period has taken active role in creating a cyber security eco-system, where tricking, cyber attacking, and cyber defensing in terms of allies or direct purposes were often observed along 9-day camp. The findings in this manner are valuable for educational outcomes and point that such educational activities can be more effective if some unexpected situations and social experiments can be included, too. Out of the planned activities, it could be better to leave some uncontrolled parameters so that such educational processes can be more effective over individuals. As the nature camp of the project was conducted in the natural environment of Egirdir, Isparta (Turkey), interaction with the real world in addition to other technologically caused interactions have enabled the project staff to get more than desired for having successful results with the project.

Obtained positive findings have encouraged the author (project leader) to think about future works more. As indicated before by Kose (2018), a second version of the project has already planned with the title 'Another Cyber Security Adventure' (with also an acronym: $B^2SGM$, as derived from the Turkish title: *'Bir Başka Siber Güvenlik Macerası'*). That second version will include a wider scenario with more technology use and just waits for funding to be realized. Additionally, more analyses regarding the unexpected cyber security eco-system were planned as additional future works. Finally, there will be alternative projects for different topics (i.e. robotics, future of Artificial Intelligence) and these projects will certainly include some spaces to open doors for unexpected experiments, which seem very critical and remarkable to be analyzed for better educational outcomes.

*NOTE: This study was supported by TÜBİTAK (The Scientific and Technological Research Council of Turkey) as 4004 – Nature Education and Science Schools project call (Project No: 118B289).*